\documentclass[prd,aps,showpacs,preprintnumbers]{revtex4}

\usepackage{graphicx,epsfig}% Include figure files
\usepackage{amssymb,amsmath}
\usepackage{bm}

\newcommand{\be}{\begin{equation}}
\newcommand{\ee}{\end{equation}}
\newcommand{\bea}{\begin{eqnarray}}
\newcommand{\eea}{\end{eqnarray}}
\newcommand{\ba}{\begin{array}}
\newcommand{\ea}{\end{array}}
\newcommand{\beqn}{\begin{eqnarray*}}
\newcommand{\eeqn}{\end{eqnarray*}}

\begin{document}

\nopagebreak

\title{Eccentricity content of binary black hole initial data}

\author{Emanuele Berti}
\email{berti@wugrav.wustl.edu}

\author{Sai Iyer}
\email{sai@physics.wustl.edu}

\author{Clifford M. Will}
\email{cmw@wuphys.wustl.edu}

\affiliation{\it McDonnell Center for the Space Sciences, 
Department of Physics, Washington University, 
St. Louis, Missouri 63130, USA}

\begin{abstract}
\noindent

Using a post-Newtonian diagnostic tool developed by Mora and Will, we
examine numerically generated quasiequilibrium initial data sets 
that have been used in recently successful numerical evolutions of
binary black holes through plunge, merger and ringdown.  We show that 
a small but significant orbital eccentricity
is required to match post-Newtonian and quasiequilibrium calculations.
If this proves to be a real eccentricity,
it could affect the fine details of the subsequent numerical
evolutions and
the predicted gravitational waveforms.

\end{abstract}

\maketitle

\section{Introduction}
\label{sec:intro}

%%%%%%%%%%%%%%%%%%%%%%%%%%%%%%%%%%%%%%%%%%%%%%%%%%%%%%%%%%%%%%%%%%%%%%%%%%%%%%%
%%%%%%%%%%%%%%%%%%%%%%%%%%%%%%%%%%%%%%%%%%%%%%%%%%%%%%%%%%%%%%%%%%%%%%%%%%%%%%%

Recent breakthroughs in numerical relativity have made it possible to
evolve Einstein's equations for binary black holes (BBH) stably for several
orbits, including the plunge, merger and ringdown phases, and to
generate intriguingly robust gravitational waveforms \cite{Pretorius,
utb06,baker06,diener06,hermann06}.  The starting
point of these evolutions is a set of initial data, obtained from the
initial-value equations of general relativity, intended to represent
two black holes in circular orbital motion; this is the expected end
product of long-term binary evolution under the circularizing and
damping effects of gravitational radiation emission. 

In earlier work
\cite{MW1,MW2} we developed an approach, based on the post-Newtonian
approximation, designed to study and elucidate the physical content of
these initial data sets, and showed that, in order to match
post-Newtonian theory with some data sets
\cite{GGB1,GGB2}, 
a small but significant orbital eccentricity was required. 
In this paper we apply this post-Newtonian diagnostic to the initial
data used in recent BBH evolutions, and find that they also require
an orbital eccentricity.  In particular we examine the 
corotating and non-spinning
initial data computed by Cook and collaborators \cite{cook04,caudill}
and the non-spinning ``puncture'' initial data of Tichy and Br\"ugmann and of
the ``Lazarus'' group \cite{lazarus,tichy}.
If this residual and unintended non-circularity is real, it may affect the
detailed structure of the numerically generated gravitational waveforms.

The plan of the paper is as follows. In
Sec.~\ref{pnd-eqs} we summarize the Post-Newtonian diagnostic
equations for BBH
derived in \cite{MW2}. In Sec.~\ref{sec:QEBH} we
apply the diagnostic to the BBH quasiequilibrium configurations of
\cite{cook04,caudill,lazarus,tichy}.
Sec.~\ref{sec:conclusion} presents conclusions.

\section{Post-Newtonian diagnostic for binary black holes}
\label{pnd-eqs}

Consider a binary system of black holes of irreducible masses $m_1$
and $m_2$, and rotational angular velocities $\omega_1$ and
$\omega_2$, with $m_{\rm irr}=m_1 + m_2$ and $\eta = m_1m_2/m_{\rm
irr}^2$ defining the total irreducible mass
and reduced mass parameter, respectively ($0 < \eta \le 1/4$).
Following \cite{MW1,MW2} we define an eccentricity $e$ and a quantity
$\zeta \equiv m_{\rm irr}/p$ related to the semi-latus rectum $p$ of the orbit,
according to:
\begin{eqnarray}
e &\equiv& \frac{ \sqrt{\Omega_p} - \sqrt{\Omega_a} } 
	{ \sqrt{\Omega_p} + \sqrt{\Omega_a} } \,,
\nonumber \\
\zeta \equiv
\frac{m_{\rm irr}}{p} &\equiv& \left ( \frac{\sqrt{m_{\rm irr}\Omega_p} +
\sqrt{m_{\rm irr}\Omega_a}}{2} \right )^{4/3} \,,
	\label{ezeta}
\end{eqnarray}
where $\Omega_p$ is the value of the orbital angular frequency $\Omega$ where
it passes through a local maximum (pericenter), and $\Omega_a$ is the
value of $\Omega$ where it passes through the {\it next} local minimum
(apocenter).  These quantities reduce exactly to their Newtonian
counterparts in the small orbital frequency (Newtonian) limit, 
and are gauge invariant through second-post
Newtonian order, among other advantages \cite{MW1,MW2}.

We want to compare with
quasiequilibrium configurations of equal-mass BH-BH binaries,
so we set $m_1=m_2$ and $\eta=1/4$. For
corotating binaries we also set ${\omega}_1={\omega}_2 \equiv \omega
=\Omega$, while for non-spinning binaries we have
$\omega_1=\omega_2 =0$.  We exploit the fact
that there exist exact formulae for the energy and spin of isolated
Kerr black holes in terms of the irreducible mass, $M=M_{\rm
irr}/[1-4(M_{\rm irr}\omega)^2]^{1/2}$, $S=4M_{\rm irr}^3
\omega/[1-4(M_{\rm irr}\omega)^2]^{1/2}$.  The total binding energy and
angular momentum of the system are then given
by
\begin{eqnarray}\label{BHBH-EJeq}
E_{\rm b} &=& E_{\rm Self} + E_{\rm Orb} + E_{\rm N,Corr} + E_{\rm Spin} \,,
\nonumber \\
J &=& S + J_{\rm Orb} + J_{\rm N,Corr} + J_{\rm Spin} \,,
\end{eqnarray}
where
\begin{subequations}
\begin{eqnarray}
E_{\rm Self} &=& m_{\rm irr} \left [ \frac{1}{2}(m_{\rm irr}\omega)^2
+ \frac{3}{8}(m_{\rm irr}\omega)^4 + \dots \right ]
\,, 
\label{Ebhself}
\\
S &=& m_{\rm irr}^3\omega \left [ 1 + \frac{1}{2}(m_{\rm irr}\omega)^2
+ \frac{3}{8}(m_{\rm irr}\omega)^4 + \dots \right ]
\,, 
\label{Jbhself}
\\
E_{\rm Orb} &=&
-\frac{1}{8}m_{\rm irr} (1-e^2)\zeta \left [ 1 - \frac{1}{48} (37-e^2)\zeta
-\frac{1}{384}(1069-718e^2+57e^4)\zeta^2
\right .
\nonumber \\
&& \left . +\left ( \frac{1}{331776} (1427365-434775e^2+110127e^4-3133e^6) 
- \frac{41 \pi^2}{384} (5-e^2)  \right ) \zeta^3 \right ] \,,
\label{Eharmdiag}
\\
J_{\rm Orb} &=&
\frac{1}{4}m_{\rm irr}^2 \frac{1}{\sqrt{\zeta}}
\left [ 1 + \frac{1}{24}(37-e^2)\zeta
+\frac{1}{384}(1069+450e^2-55e^4) \zeta^2
\right .
\nonumber \\
&& \left .
-\left (\frac{1}{82944}(285473-271419e^2-93e^4-713e^6)
-\frac{41 \pi^2}{96}(1+e^2) \right ) \zeta^3 \right ]
\,,
\label{Jharmdiag}
\\
E_{\rm N,Corr} &=& -\frac{5}{48}m_{\rm irr}(1-e^2)(m_{\rm irr}\omega)^2
\zeta \,,
\\
J_{\rm N,Corr} &=& \frac{5}{24}m_{\rm irr}^2 (m_{\rm irr}\omega)^2
/\sqrt{\zeta} \,,
\\
E_{\rm Spin} &=& -\frac{1}{12} m_{\rm irr}(1-e^2)(7-2e^2)(m_{\rm irr}\omega)
\zeta^{5/2} \,,
\\
J_{\rm Spin} &=& -\frac{5}{24}m_{\rm irr}^2 (7+e^2) (m_{\rm irr}\omega) \zeta \,.
\end{eqnarray}
\end{subequations}
The orbital (``Orb'') contributions are expressed in the ADM 
(Arnowitt-Deser-Misner) gauge and
are valid to third post-Newtonian (3PN) order.  
In Eqs. (\ref{Ebhself}) and (\ref{Jbhself}), we have expanded the Kerr
formulae for $M$ and $S$ in powers of $m_{\rm irr}\omega$, assumed to
be small compared to unity, keeping as many terms as
needed to reach a precision comparable to our 3PN orbital formulae, and have
subtracted $m_{\rm irr}$ in order to obtain the binding energy. 
The ``N,Corr'' terms come from converting
the individual {\em total} masses that appear in the Newtonian orbital energy to
irreducible masses and their corrections due to spin, and the ``Spin''
terms are spin-orbit effects.
For black hole binaries, tidal and spin-spin effects can be shown to be
negligible \cite{MW2}.  
To obtain $E_b$ and $J$ at a turning point as
functions of $\Omega$, we substitute $\zeta = (m_{\rm
irr}\Omega_a)^{2/3}/(1-e)^{4/3}$ or $\zeta = (m_{\rm
irr}\Omega_p)^{2/3}/(1+e)^{4/3}$ for apocenter or pericenter,
respectively.  
When $E_b$, $J$, $\omega$ and $\Omega$ are suitably
scaled by $m_{\rm irr}$, there
remains only one free parameter, the eccentricity of the orbit, 
This approach was used in \cite{MW1} to compare
with the numerical quasiequilibrium solutions of Grandcl\'ement
{\it et al.} \cite{GGB2},  
and it was
found that a substantially better fit to the numerical
data was obtained for non-zero values of $e$, of the order of $0.03$,
with the system at apocenter, than for $e=0$.  We now apply this
diagnostic to other data sets that have recently played an important
role in BBH evolutions.

\begin{figure*}[htb]
\begin{center}
\begin{tabular}{cc}
\epsfig{file=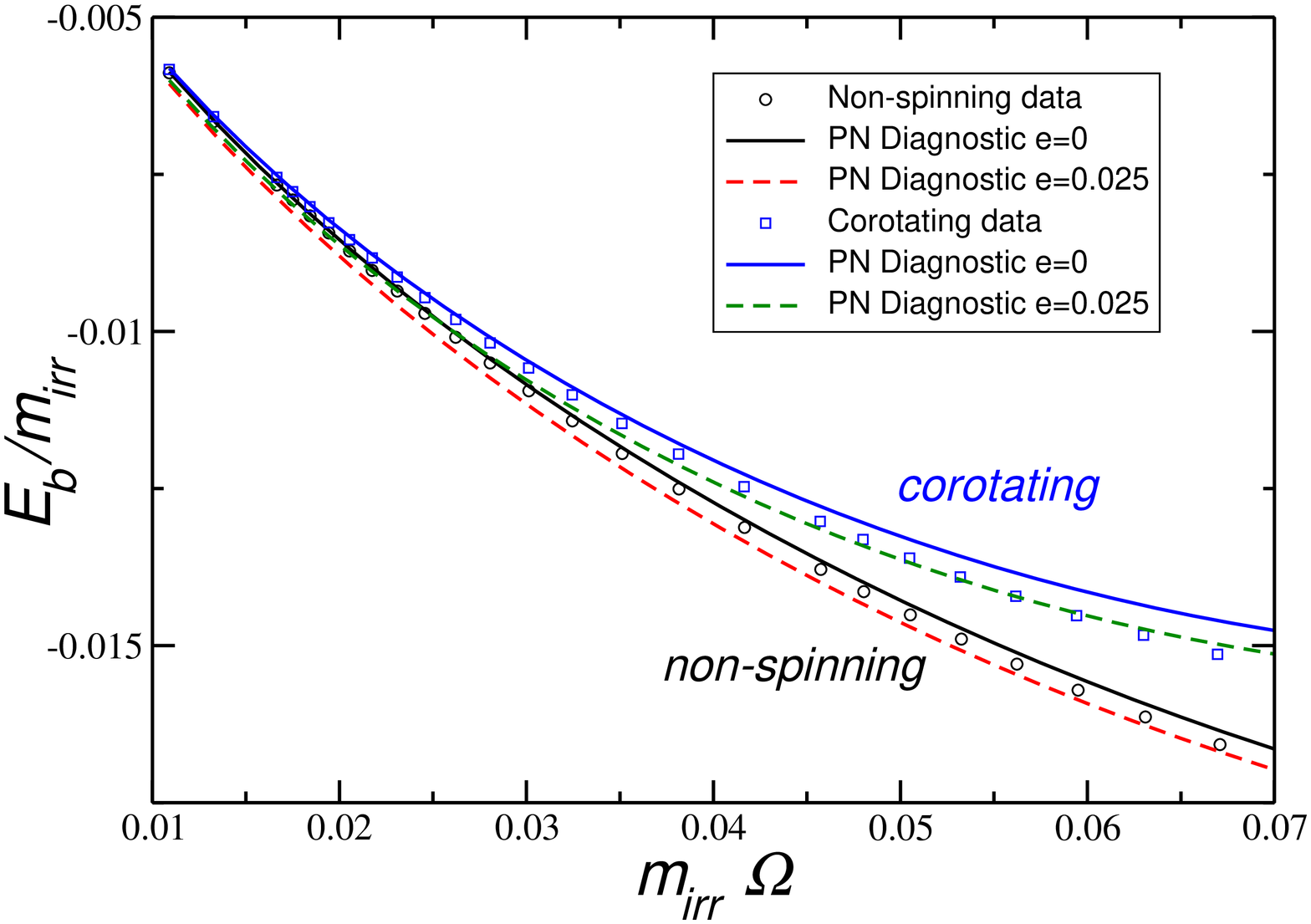,width=6.0cm,angle=-90} &
\epsfig{file=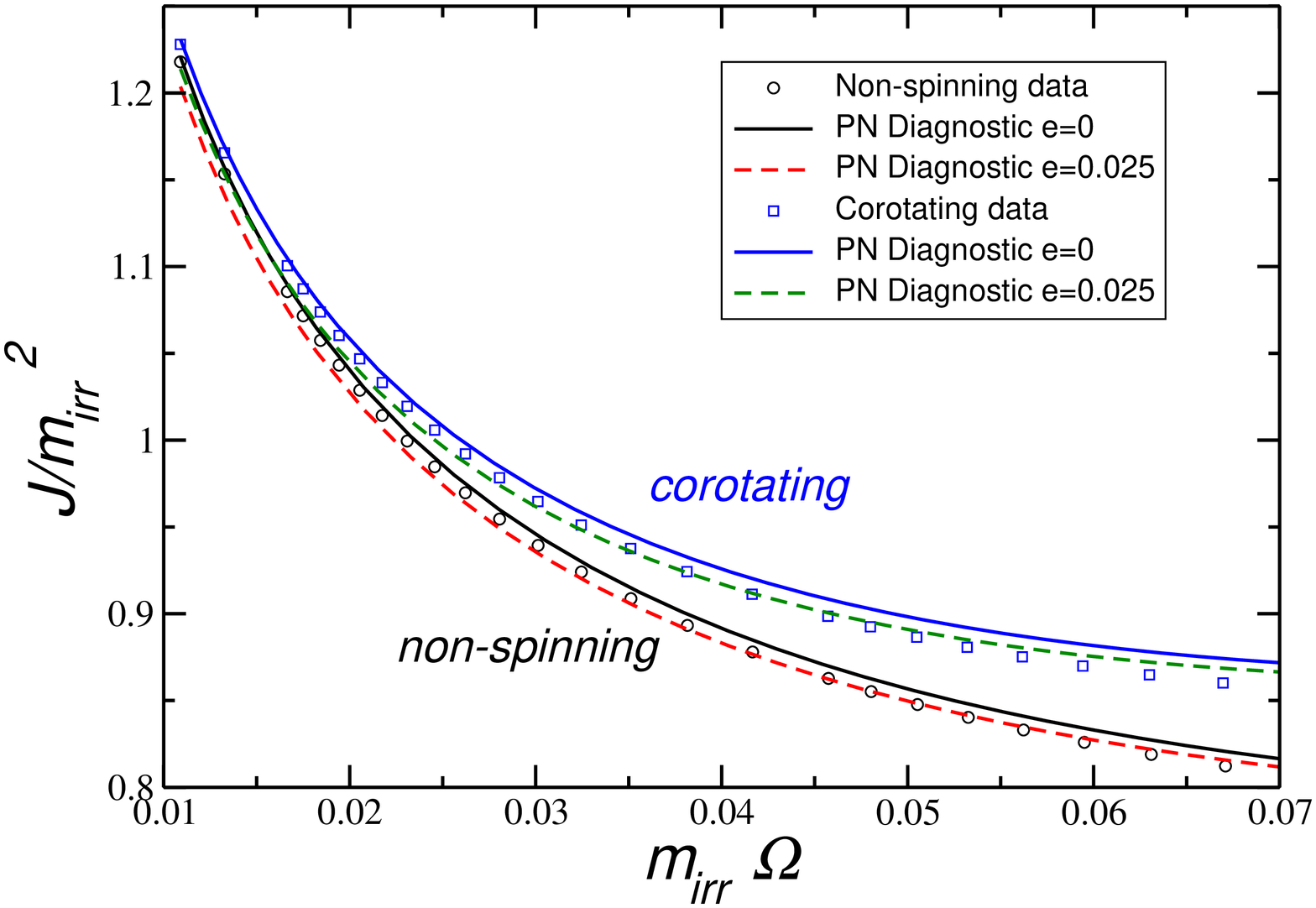,width=6.0cm,angle=-90} 
\end{tabular}
\caption{Binding energy (left) and angular momentum (right) for the
corotating and non-spinning BH-BH initial data from
\cite{caudill}. Circles (squares) are non-spinning (corotating) data
from \cite{caudill}, the solid lines are circular PN diagnostics with
$e=0$, the dashed lines are eccentric PN diagnostics with $e=0.025$.
\label{Caudill}}
\end{center}
\end{figure*}
\begin{figure*}[htb]
\begin{center}
\begin{tabular}{cc}
\epsfig{file=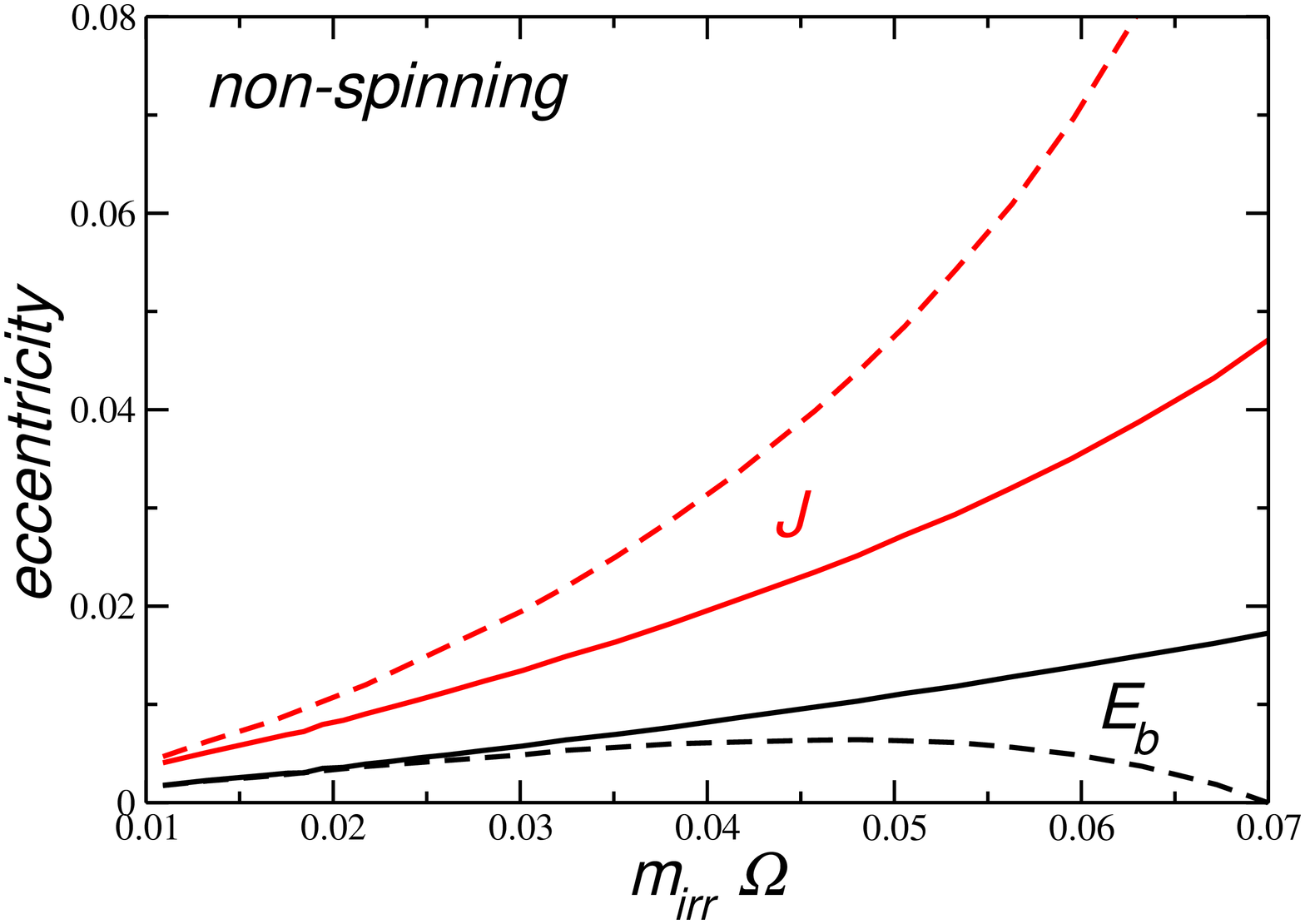,width=6.0cm,angle=-90} &
\epsfig{file=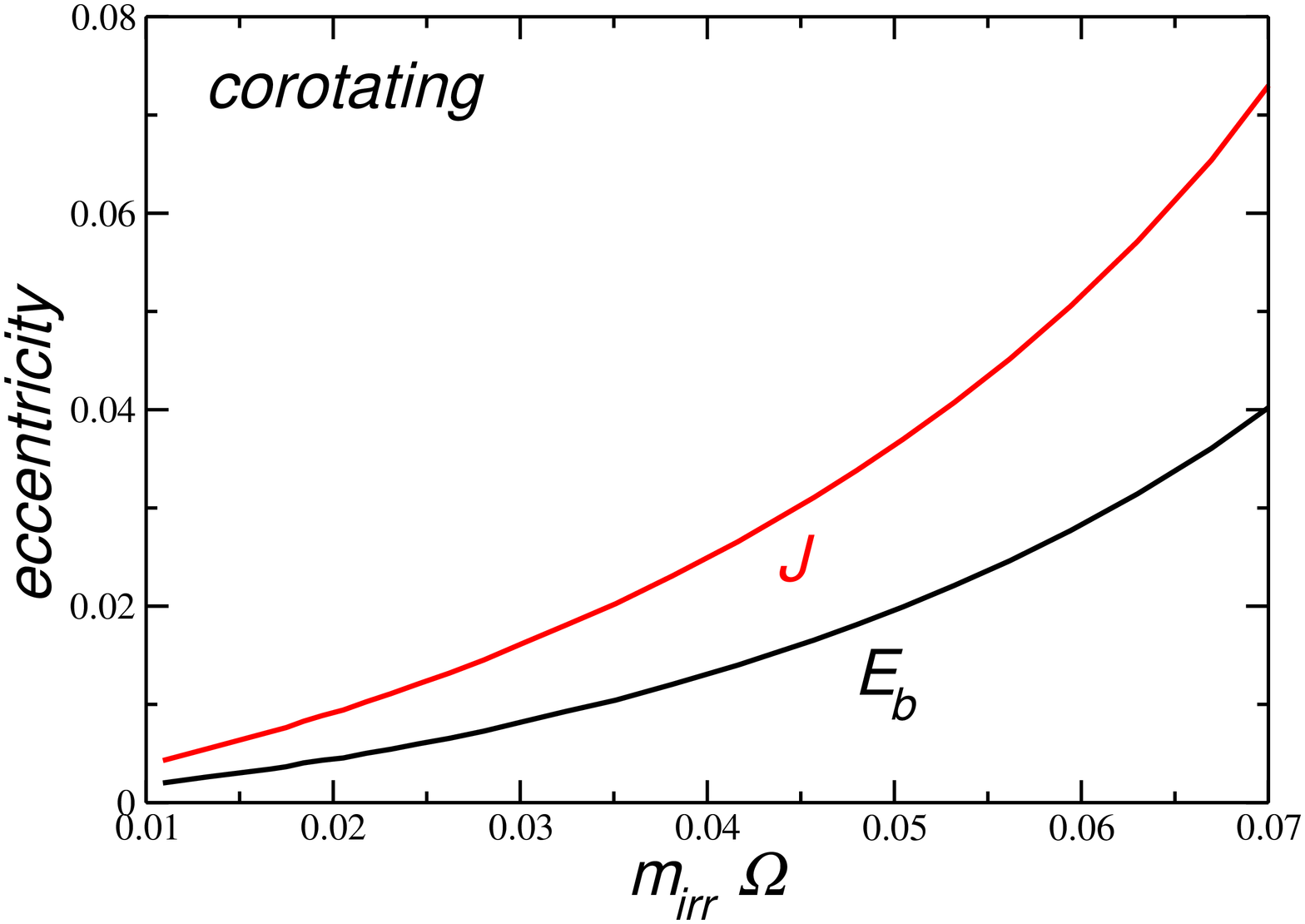,width=6.0cm,angle=-90} 
\end{tabular}
\caption{Eccentricity required to match data points
for non-spinning (left) and corotating
data (right).  Black lines are solutions for the energy, red lines are 
solutions for the angular momentum. In the non-spinning case, continuous
lines refer to the Caudill {\it et al.} data \cite{caudill},
dashed lines to the earlier Cook-Pfeiffer data \cite{cook04}.
\label{Caudill2}}
\end{center}
\end{figure*}
\section{Diagnosis of BBH initial data sets}
\label{sec:QEBH}

Cook and collaborators have developed initial data sets for
quasiequilibrium BBH, allowing for both corotation and zero spin, in
a series of papers
\cite{cook93,cook94,cook94b,cook02,cook02b,cook04,caudill}.  Using the
thin-sandwich approach, combined with ``excision'' boundary conditions
for the black holes adapted for treating spin,
they considered systems possessing a ``helical Killing vector'',
$\partial/\partial t + \Omega \partial/\partial \phi$, meant to
represent a circular orbit, one that is stationary in a frame
rotating with angular velocity $\Omega$.  
Additionally, they impose the condition that the Komar
mass, a mass defined for stationary systems, equal the ADM
mass, an invariant mass measured at spatial
infinity.  It is believed that this condition helps ensure that the
orbit is truly circular.  
We apply
our diagnostic to two data sets, taken from Refs. \cite{cook04} and
\cite{caudill}, respectively.  For non-spinning BH, the second
data set used a more
accurate prescription for setting the BH spins to zero; in the earlier
data, the black holes were not truly nonrotating.
We take the data from Tables IV (corotating)
and V (non-spinning) of \cite{cook04} and of \cite{caudill}, and plot
$E_b/m_{\rm irr}$ and $J/m_{\rm irr}^2$ vs. $m_{\rm irr}\Omega$.
Figure \ref{Caudill} shows the comparison between the data of Caudill
{\it et al.}  \cite{caudill} and our diagnostic, plotted for $e=0$ and
for $e=0.025$ (with our definitions, positive values of $e$ correspond
to the system being at apocenter).  Figure \ref{Caudill2} shows the
eccentricity required to match each data point from both
\cite{cook04} and \cite{caudill}.  In the 
improved data set of Caudill {\em et al.} for the non-spinning case, the
apparent eccentricity in the fits to $J$ is reduced, and
the functional behavior of $e$ with $m_{\rm irr}\Omega$ is now the same
(monotonically increasing) in both the non-spinning and corotating
cases.  For the
corotating case, there is essentially no difference in the fits
between the two data sets.  Furthermore, as in earlier comparisons
\cite{MW1}, there is a systematic difference between the eccentricity
required to match the binding energy and that required to match the
angular momentum.

\begin{figure*}[htb]
\begin{center}
\epsfig{file=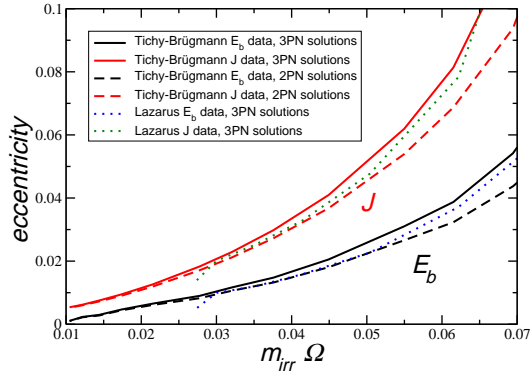,width=6.0cm,angle=-90} 
\caption{Eccentricity solutions for 
puncture initial data.  Solid and dashed lines are solutions for the
Tichy-Br\"ugmann puncture data using 3PN and 2PN diagnostics,
respectively.  Dotted lines are from 3PN solutions for the Lazarus
puncture data.
\label{tblaz}}
\end{center}
\end{figure*}

Another approach to initial quasiequilibrium data for BBH evolutions
is the ``puncture'' method, in which the conformal factor of the
conformally flat spatial slices is written in terms of a
Newtonian-like potential $m_A/|{\bf x}-{\bf x}_A|$ (a ``puncture'')
for each body.  This approach can also be made to incorporate the
helical Killing vector and Komar-ADM mass equality assumptions
\cite{TBL03}.  Tichy and Br\"ugmann \cite{tichy} and Baker {\em et
al.} \cite{lazarus} (the ``Lazarus'' project) have used this approach
to generate initial data for non-spinning binary black holes in
quasi-circular orbits.  In Figure \ref{tblaz} we show the 
eccentricity required to match these data sets; again the eccentricity
is small but significant, and again different between $E_b$ and $J$.
We also show explicitly the eccentricity required to match the
Tichy-Br\"ugmann data using a 2PN diagnostic, obtained by truncating
the 3PN terms in the orbital expressions; the differences are
comparable to the differences between the two numerical data sets, and
are not large enough to account for the difference between the $E_b$
and $J$ curves. Table \ref{tab:eccfit} lists the coefficients of a
cubic fitting function to the eccentricity required to match
$E_b$ and $J$ for different initial data sets.

\begin{table}[hbt]
\centering
\caption{Fit of the eccentricity for the corotating (corot)
and non-spinning (nospin) data by Caudill {\em et al.}, for the
Tichy-Br\"ugmann puncture data (TB) and for the Lazarus data
(Lazarus). The integer $N$ is the number of data points used for the
fit. We carry out a least-squares 
fit by a cubic polynomial $e=\sum_{k=0}^3 e_k(m_{\rm irr}\Omega)^k$.
%considering data in the range $m_{\rm irr}\Omega=[0.0109,0.0671]$ for
%the Caudill {\em et al.} data, $m_{\rm irr}\Omega=[0.0105,0.0694]$ for the
%Tichy-Br\"ugmann data and $m_{\rm irr}\Omega=[0.0300,0.0622]$ for the
%Lazarus data.
$\Delta e^{\rm max}={\rm max}[(e-e^{\rm num})/e^{\rm num}]$ is the
maximum percentage error of the fit with respect to the numerical data
$e^{\rm num}$.}
\vskip 12pt
\begin{tabular}{@{}ccccccc@{}}
\hline
\hline
\multicolumn{7}{c}{Fit of $E_b$}\\
\hline
\hline
&$N$ &$10^3\times e_0$ &$e_1$ &$e_2$ &$e_3$ &$\Delta e^{\rm max}$\\
\hline
corot   &25 &-0.73418894 &0.23870832 &-0.37207668 &74.523304 &-3.758\\
nospin   &25 &-0.11438527 &0.15748053 & 1.2258277  &0.83081005& 5.751\\
TB      &12 &-5.4145414  &0.75157074 &-14.093449  &224.07983 &-11.64\\
Lazarus &5  &-2.1760344  &0.60151690 &-13.280282  &221.40238 & 1.244\\
\hline
\hline
\multicolumn{7}{c}{Fit of $J$}\\
\hline
\hline
&$N$ &$10^3\times e_0$ &$e_1$ &$e_2$ &$e_3$ &$\Delta e^{\rm max}$\\
\hline
corot   &25 &-1.8164595  &0.56024586 &-3.2850593  &147.08294 &-4.452\\
nospin   &25 &-0.99496055 &0.45656217 &-1.0622433  &60.466300 &-3.088\\
TB      &12 &-9.4590074  &1.6687984  &-40.943406  &624.96446 &-20.98\\
Lazarus &5  &-16.332921  &2.0037212  &-43.661018  &577.56427 & 1.161\\
\hline
\hline
\end{tabular}
\label{tab:eccfit}
\end{table}

\section{Conclusions}
\label{sec:conclusion}

We have shown, using a post-Newtonian diagnostic tool, that initial
data sets for binary black hole mergers may actually represent slightly
eccentric orbits.  
Several remarks are called for.  First, there is evidence from the
dynamical evolutions of some of these initial data sets that the orbits {\em
are} slightly eccentric.   For example, in the recent evolution of
several BBH orbits through merger and ringdown by Baker {\em et al.}
\cite{baker06} using the Lazarus initial data, an oscillatory behavior 
of the separation
of the black holes can be seen in their Figure 9.  Similar
oscillations were seen in the binary neutron star evolutions of
\cite{miller04a,duez03}, although the results there were very
sensitive to grid resolution and size of the computational domain. 

On the other hand, we continue to be puzzled by the difference in
values of $e$ inferred from fits to $E_b$ and $J$.  This difference
was also seen
in fits of the diagnostic to data from the Meudon group \cite{MW1},  
and could be cited as a defect of the PN approximation.  However, this
difference occurs systematically even at the smallest values of
$m_{\rm irr}\Omega$, where relativistic corrections are quite small.

We want to emphasize that the eccentricity we are discussing here
is not related to the
mismatch between an initial quasicircular orbit (with $\dot r =
\ddot r =0$ by
construction) and the reality of a pre-existing inspiral (with $\dot r
\approx -16(m/r)^3/5$), since the initial data sets know nothing about
radiation reaction.  That eccentricity, which would be induced on
an evolution from a perfectly circular initial orbit, has been
discussed in detail by Miller \cite{miller04b}.  Depending on the
starting point of the evolution, the induced eccentricity from this
effect could be as
large as 0.03.

Irrespective of the origin of the eccentricity, Miller pointed out
that the result could be a
substantial decrease in  detection
signal-to-noise when a numerically
generated, eccentric waveform template is matched against a ``true''
waveform generated by a quasicircular inspiral of a real BBH (see, for
example Figures 7 and 9 of \cite{miller04b}).  

If eccentricity {\em is} an issue and cannot be removed or reduced by tuning
the initial data sets, one could ask whether
it could be damped away naturally by numerically
evolving several orbits leading up to the
onset of plunge, around $m_{\rm
irr}\Omega \sim 0.1$.  Using Eqs. (2.34) of \cite{MW2}, which give
the evolution of our orbit elements $e$ and $\zeta$ under radiation
reaction, it is straightforward to show, at 2.5PN order and in the
small eccentricity limit, that the number of orbits $N$ required to
reduce the eccentricity by a factor $X = e_f/e_i$ by the time the orbit reaches
a final angular velocity $\Omega_f$ is
given by $N =  (X^{-30/19}-1)/64\pi\eta(m_{\rm irr}\Omega_f)^{5/3}$.
For $\eta=1/4$ and $m_{\rm irr}\Omega_f = 0.1$, this gives 34 orbits
for a reduction by 1/10, and 11 orbits for a reduction by 
1/5.  Suppressing eccentricity this way is likely to be a challenge 
without additional breakthroughs in
numerical relativity.

\acknowledgments

This work was supported in part by the National Science Foundation
under grant number PHY 03-53180.
CMW is grateful to the Groupe Gravitation Relativiste et Cosmologie
(GR$\varepsilon$CO) of the Institut d'Astrophysique de Paris for its
hospitality during the completion of this work.  We thank Greg Cook
and the participants of the Astrophysical Applications of Numerical
Relativity Workshop (Guanajuato, Mexico) for discussions.

\bibliography{biw-bh}
\end{document}